\documentclass[conference]{IEEEtran}
\usepackage{blindtext, graphicx}
\usepackage{amsmath} 
\usepackage{amssymb}  
\usepackage{cite}
\usepackage{subcaption}
\usepackage{amsfonts}
\usepackage{algorithmic}
\usepackage{textcomp}
\usepackage{graphicx}
\usepackage{hyperref}
\usepackage{color}

\usepackage [english]{babel}
\usepackage [autostyle, english = american]{csquotes}
\MakeOuterQuote{"}

\ifCLASSINFOpdf
\else
\fi

\begin{document}
%
\title{Infrastructure Enabled Autonomy: A Distributed Intelligence Architecture for Autonomous Vehicles }

\author{\IEEEauthorblockN{Swaminathan Gopalswamy ~~~~~~~ Sivakumar Rathinam}
\IEEEauthorblockA{Connected Autonomous Safe Transportation (CAST) Program \\
Department of Mechanical Engineering,  Texas A\&M University, 
College Station, TX
}
}

\maketitle
%

\begin{abstract}
Multiple studies have illustrated the potential for dramatic societal, environmental and economic benefits from significant penetration of autonomous driving. However, all the current approaches to autonomous driving require the automotive manufacturers to shoulder the primary responsibility and liability associated with replacing human perception and decision making with automation, potentially slowing the penetration of autonomous vehicles, and consequently slowing the realization of the societal benefits of autonomous vehicles. 
We propose here a new approach to autonomous driving that will re-balance the responsibility and liabilities associated with autonomous driving between traditional automotive manufacturers, infrastructure players, and third-party players. Our proposed distributed intelligence architecture leverages the significant advancements in connectivity and edge computing in the recent decades to partition the driving functions between the vehicle, edge computers on the road side, and specialized third-party computers that reside in the vehicle.  Infrastructure  becomes a critical enabler for autonomy. 
With this Infrastructure Enabled Autonomy (IEA) concept, the traditional automotive manufacturers will only need to shoulder responsibility and liability comparable to what they already do today, and the infrastructure and third-party players will share the added responsibility and liabilities associated with autonomous functionalities. We propose a Bayesian Network Model based framework for assessing the risk benefits of such a distributed intelligence architecture. 
An additional benefit of the proposed architecture is that it enables "autonomy as a service" while still allowing for private ownership of automobiles.
\end{abstract}

%
\IEEEpeerreviewmaketitle

\section{Introduction}
Transportation systems and associated mobility are at the cusp of a tectonic shift.  Human beings have traditionally been in the driver’s seat of automobiles and that is beginning to change with the emergence of various automation capabilities in automobiles. This shift is heralded as potentially providing huge benefits to the society at large. Studies predict that with a 50\% penetration, autonomous vehicles will result in 9,600 lives saved per year, 1.9 million fewer crashes, \$50 billion in economic savings, 1.6 billion hours saved through less time traveled, and 224 million less gallons of fuel consumed \cite{EnoReport}. 

The primary rationale for why autonomous driving will improve safety is the premise that automobile technologies will be mature enough that they will be inherently safe. Human errors cause the majority of all traffic accidents \cite{NHTSASurvey}, and by automating human decision making, we could improve overall safety.  

The auto industry has evolved over decades, with safety as a primary driving factor in the design and development. While the physical components of the modern automobile have become quite safe and reliable (failures of engines, transmissions or other such systems are quite rare), the safety-focus has been challenged by the rapid growth in both scope and complexity of embedded software functionality in cars. The number of software related recalls are growing exponentially \cite{JDPowerSurvey}.  Such safety concerns are exacerbated for autonomous vehicles, where we are trying to replace human decision making with algorithms. The use of machine learning for both perception and decision making brings in an inherent non-determinism to the system performance making it nearly extremely difficult, if not impossible, to assert performance safety of the autonomous vehicles. (e.g. \cite{Koopman}.

Thus, for autonomous vehicles, the automotive OEM becomes saddled with both the responsibility and liabilities associated with the traditional capabilities of the vehicle, but also those associated with functions that human beings routinely perform.  In section II below, we look at this  distribution in more detail, and propose a new architecture  that effectively reduces this liability to the automotive OEMs through a re-balancing with the infrastructure.  Section III describes the new proposed concept in greater detail. Section IV provides a mathematical framework for analyzing the reduction in risks and demonstrates this through a numerical example. Section V provides some conclusions including the value proposition of the concept and open research themes for further  consideration.

\section{Distribution of Liability and Responsibility in Personal Automotive Transportation}

\subsection{Distribution of responsibility and liability for modern non-autonomous cars}

Using a high level functionally decomposition of the modern automobile with its driver and connectivity, we can identify a distribution of responsibilities as illustrated in the schematic of Fig.\ref{fig:CurrentDist} below.

\begin{figure*}[htb]
\centering{}
\includegraphics[scale=0.5]{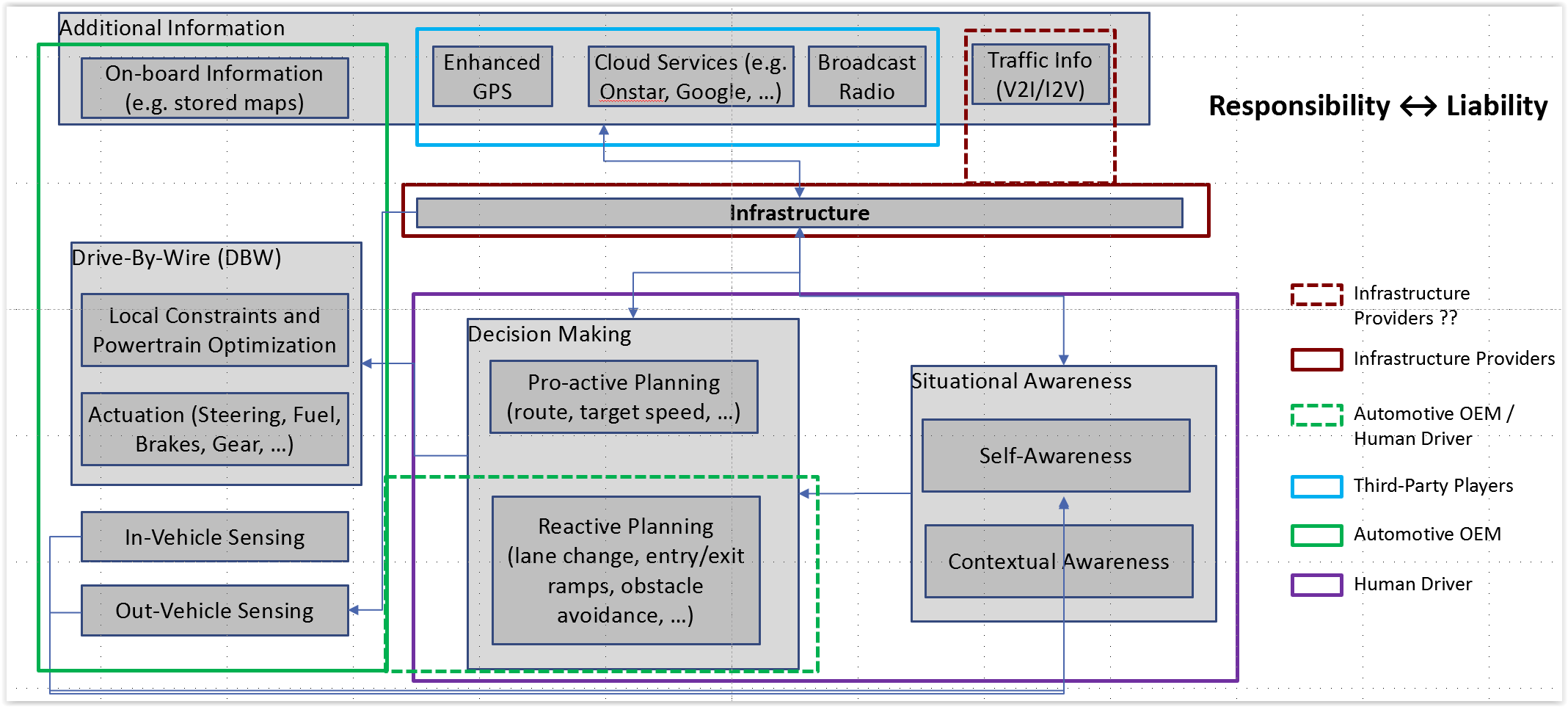}
\caption{Current distribution of driving responsibilities (non-autonomous)}
\label{fig:CurrentDist}
\end{figure*}

The automotive OEM has a clear responsibility to provide the following functionalities:

\begin{itemize}
    \item Primary driving (or powertrain) functions: Namely to provide the torque, power and steering needed to drive the car. Most modern cars are “drive-by-wire (DBW) enabled, $i.e.$, the inputs from the driver (such as accelerator pedal, brake pedal or steering) are converted to appropriate actuation signals to the lower level actuators. The conversion comprehends any internal constraints arising from the physics of the system as well as those posed by usability and driver comfort.
    \item Display of diagnostic and other sensor information about the vehicle to the driver. This would include direct information such as the vehicle and engine speed as well as more nuanced information such as average fuel economy or available range, or diagnostic information such as the engine needs to be serviced shortly.
    \item Display of information outside of the vehicle. This would include passive devices such as rear-view and side-view mirrors, to more active devices such as blind-spot detection or distance to obstacles. 
\end{itemize}

The automotive OEM may source some of the capabilities from suppliers, however retains the responsibility and liability. The automaker may facilitate availability of information from outside by leveraging external infrastructure such as Enhanced Global Positioning System (GPS), cloud based services such as OnStar, Google, etc., broadcast radio, etc. The automaker might also facilitate delivery of some traffic-specific information through such infrastructure. The infrastructure is managed by infrastructure operators. The information is then used by the driver to generate situational awareness of the car and its surroundings and make appropriate decisions while driving. Situational awareness will include an understanding of the car’s own driving status (such as its velocity, acceleration, location, etc.) – called the self-awareness, and an understanding of the car’s surroundings (objects near by, their state of motion, traffic signals and other road signs, etc) – called the contextual-awareness.  The automaker OEM assumes competent driving. The driver remains responsible and liable for their driving.

The modern automobile also has some support functions from the OEMs to support some driver actions (such as lane change, obstacle avoidance, etc.). When the OEM provides such functionalities, clearly the OEM has responsibility and liability for those. Correspondingly these functions are being deployed very cautiously by the automakers. These functionalities are just “Advance Driver Assistance Systems”, and the driver remains ready to take back complete control at any time while driving.

\subsection{Distribution of responsibility and liability per the current paradigm for Autonomous Vehicles}

\begin{figure*}[htb]
\centering{}
\includegraphics[scale=0.5]{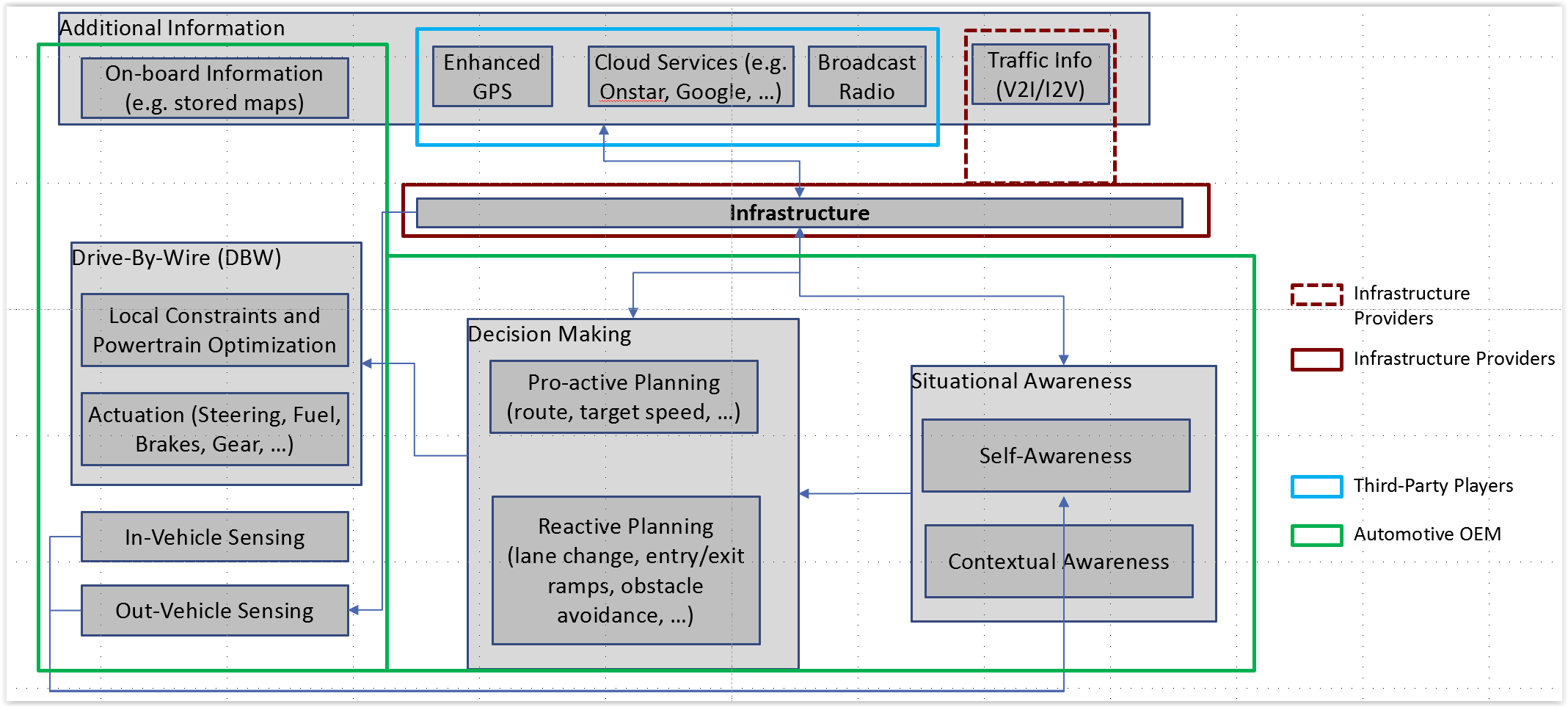}
\caption{Distribution of driving responsibility in autonomous vehicles}
\label{fig:AVDist}
\end{figure*}

The distribution of responsibility changes dramatically once we consider autonomous vehicles, especially with autonomy levels of 3 or higher. Fig.\ref{fig:AVDist} shows how the responsibility of the OEM increases to include Situational Awareness synthesis and decision making. (This is all the time for Level 5, while intermittent for other levels of automation). 

Situational awareness synthesis has made tremendous strides in recent years leveraging advances in Machine Learning, Vision Processing, and sensor technologies such as in LIDARs and RADARs,  and there is a high confidence amongst that this will be a solved problem in the near future, e.g. \cite{MobilEyePaper}. However, there continues to be a note of caution from several software safety experts (\cite{Koopman}).

Decision making continues to be a big open question because it is no longer a matter of identifying and classifying the physical world (as was in Situation Awareness synthesis), but it is a question of identifying and classifying "human intent" (of other drivers and pedestrians). While many novel methods are being applied to tackle this part of the driving functionality, (e.g. \cite{RezaLangari}), the problem fundamentally remains non-deterministic. Correspondingly the uncertainty related to the performance, and the risks and liabilities remain. Despite the very many exciting announcements about introduction of autonomous vehicles, there have also been voices of caution (e.g. \cite{GillPratt} ).

\subsection{ Proposed Paradigm-shift in the distribution of responsibilities and liabilities}

We propose a new paradigm for autonomous vehicle driving: The OEMs shall take direct responsibility (and liability) for the core capabilities related to driving (DBW and vehicle-level sensing). But they will not take direct responsibility for decision making or generating the situational awareness. 

The situational awareness will be generated through sensors that are embedded in the infrastructure. Thus, the responsibility (and liability) for situational awareness is shifted to the infrastructure operators.

Decision making is provided by yet another third party that takes the situational awareness information coming from the infrastructure operators (leveraging the connectedness of the infrastructure), and uses standardized Application Programming Interfaces  (APIs) to interface with the DBW capabilities of the OEM to drive the cars autonomously.
 
\begin{figure*}[htb]
\centering{}
\includegraphics[scale=0.5]{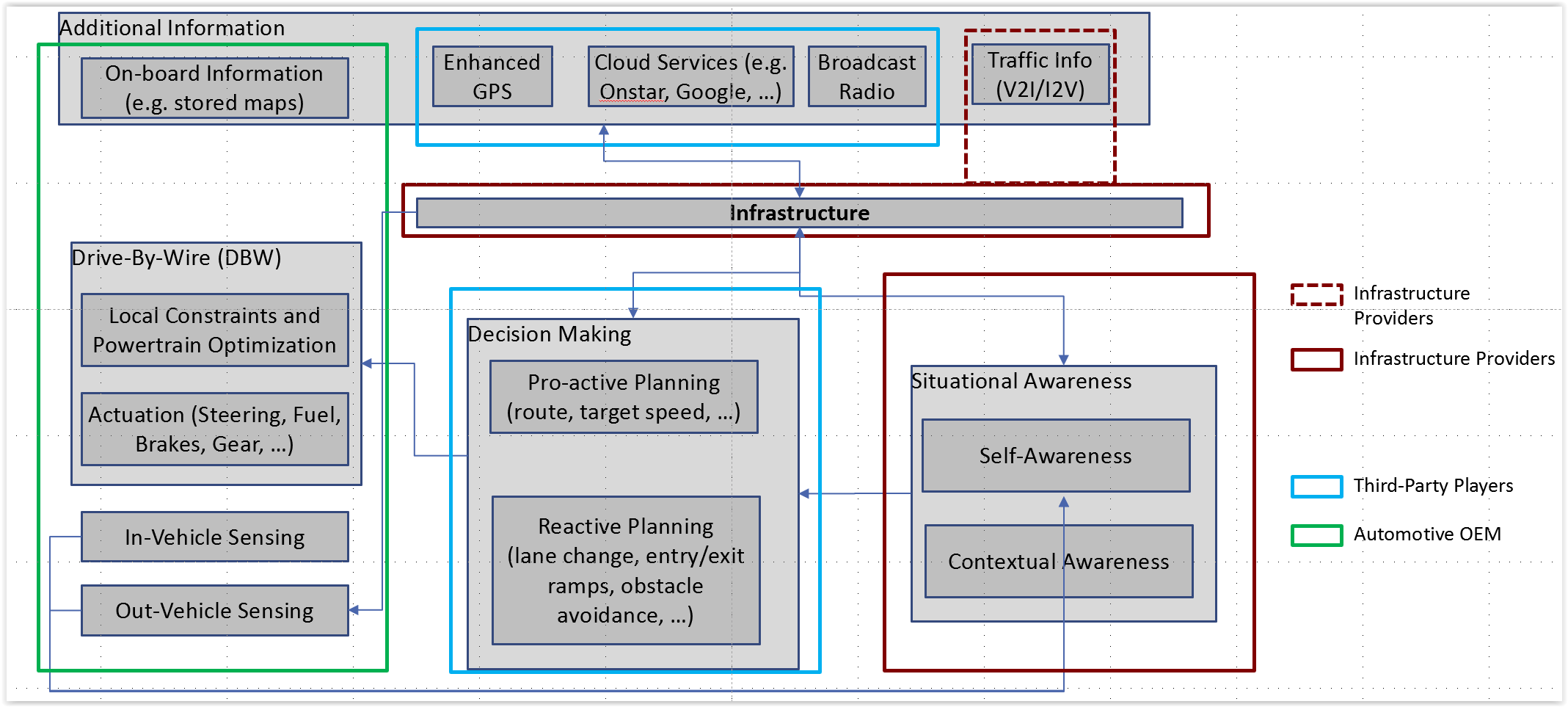}
\caption{Distribution of driving responsibility in the Infrastructure Enabled Autonomy (IEA) Concept}
\label{fig:IEADist}
\end{figure*}

This distribution is captured in Fig.\ref{fig:IEADist}. The new redistribution of the responsibilities will suddenly create opportunities for deployment hitherto not possible. The OEMs will continue to build on their core competencies. They could partner with third parties to provide decision making algorithms, with the business participation commensurate with the liability they are willing to accept. Traditional infrastructure operators (such as toll booth operators or cellular phone operators) can now take on value-added business, bringing a much needed infusion of new business to their nearly commoditized business models.

\section{The New proposed Paradigm: Infrastructure Enabled Autonomy (IEA)}

\subsection{Concept Overview}

\begin{figure*}[htb]
\centering{}
\includegraphics[scale=0.5]{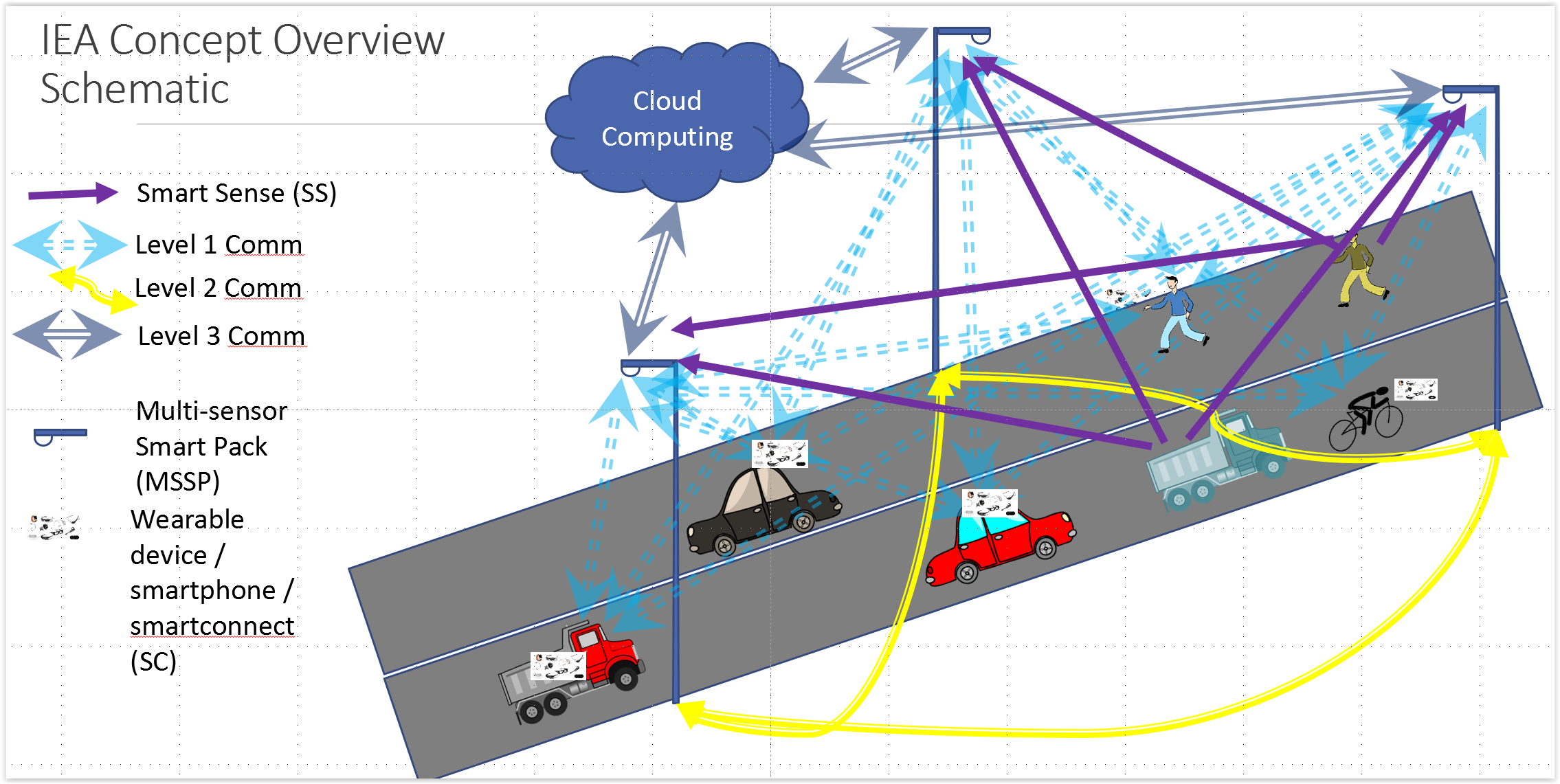}
\caption{Sample Overview Schematic of the IEA Concept }
\label{fig:IEAConceptDiag}
\end{figure*}

Fig.\ref{fig:IEAConceptDiag} is a pictorial representation of the IEA concept. IEA will be deployed on special traffic corridors that we call Special Infrastructure Enabled Traffic Corridors (SIETC). Figure 5 shows one such corridor with mixed traffic consisting of automated and manually driven vehicles, as well as other traffic such as pedestrian and bicyclists. We will call the vehicles automated through the IEA concept as the IEA vehicles.

Road-Side-Units (RSUs) on the infrastructure will be fitted with special devices that we will call Multi-Sensor-Smart-Packs (MSSPs). The MSSPs will monitor the SIETC and generate situatioal awareness (SA) information that will be transmitted using wireless means to special devices that we will call SmartConnects (SCs). 

The SCs typically reside in the IEA vehicle, and interface with the DBW capabilities of the car and provide the commands to drive autonomously. The SCs could also be deployed on manually driven vehicles, as well as individual passengers (through smart-phone-type devices). In this case, the SCs will use the SA information to provide guidance and warnings to the users.

As the vehicle travels through the SIETC, special hand-shake protocols will be used by SCs to engage with the different MSSPs along the way.

The IEA vehicles will normally be driven by a human driver. When the driver enters an SIETC, there will be an electronic engagement with the corridor, wherein the driver could choose to be driven autonomously. If the driver desires to do so, there is an appropriate handshake with the corridor, and the vehicle is driven autonomously as described above. When it it time for the driver to take control back, depending on the alertness of the driver, either the driver take control back, or the vehicle is parked in designated take-over spot. (The SIETCs will be defined to facilitate such a deployment.) A typical deployment scenario that could be envisioned with the IEA concept is shown in Fig.\ref{fig:IEADeployent} below.

\begin{figure*}[htb]
\centering{}
\includegraphics[scale=0.5]{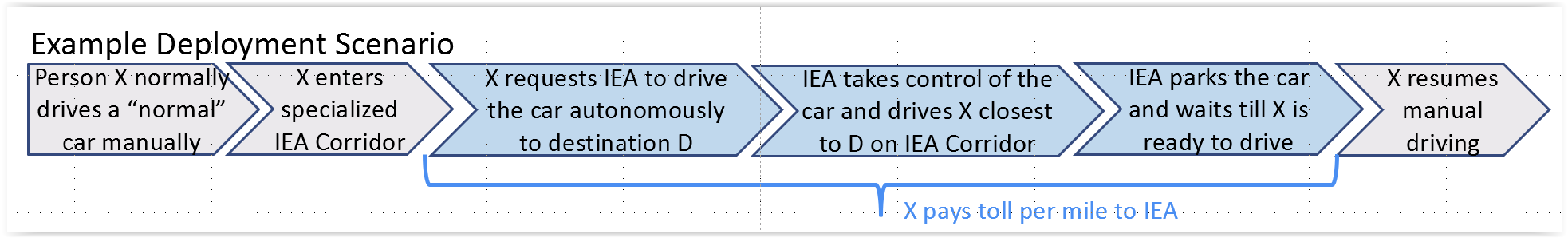}
\caption{Example Deployment Scenario with IEA}
\label{fig:IEADeployent}
\end{figure*}

\subsection{Functional Overview of the IEA}

\begin{figure}[htb]
\centering{}
\includegraphics[scale=0.8]{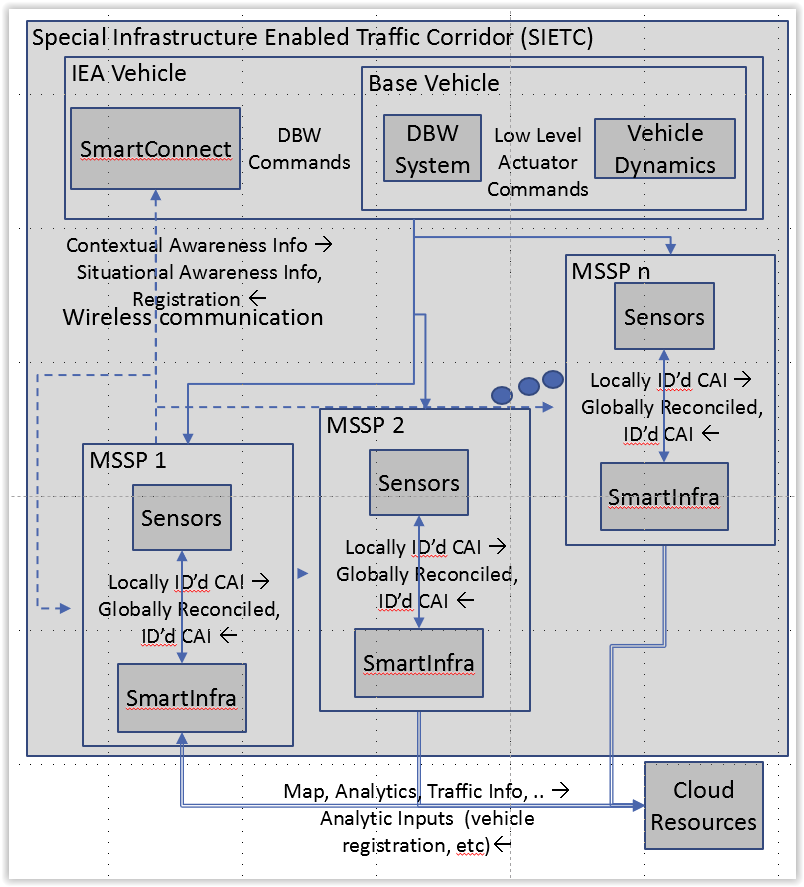}
\caption{A Functional Schematic of the IEA Concept }
\label{fig:IEAFuncSchematicVert}
\end{figure}

Fig.\ref{fig:IEAFuncSchematicVert} shows a functional architecture of the IEA concept. We will describe the major components of this IEA architecture below.

\subsection{Special Infrastructure Enabled Traffic Corridors (SIETC)}

From a business viability perspective, IEA will generate maximum value and so deployed where  traffic density is high and when the drivers are not traveling for fun. This would typically be urban commuter traffic. These roads and streets-(or designated lanes within them) will have been fitted with MSSPs, and will be connectivity-enabled. 

The SIETCs will be operated by infrastructure operators, similar to toll-booth operators. When a vehicle uses the SIETC, depending on the services that they receive from the SIETC, they will be charged by the infrastructure operator based on usage.

\subsection{Connectivity within in a SIETC}

The SIETC will have connectivity technologies that will server three levels of communication:

\begin{itemize}
    \item {\it Level 1 Communication:} This refers to communication from an MSSP to moving devices in the “neighborhood” of the MSSP. The key requirement is that this communication be wireless. The required range of communication is relatively small. Examples would include DSRC, Wifi, cellular, 5G, etc.
    \item {\it Level 2 Communication:} This refers to communication between neighboring MSSPs. This is expected to be dedicated very high speed communication, such as fiber optics.
    \item {\it Level 3 Communication:} This refers to communication between the MSSPs and a cloud-based computing capability that provides support for perception, classification, etc.
\end{itemize}

\subsection{Multiple-Sensor Smart Packs (MSSPs)}

The MSSPs will consist of (i) multi-sensor packs that monitor the road, (ii) computers that process the sensed information and generate the SA information,  (iii) wireless connectivity to transmit the SA to Smart-Connects (SCs), and (iv) appropriate power supply to support its operations.

A sensor pack would contain one or more sensors of different types (such as LIDAR, RADAR, optical camera, thermal camera, etc.). Some of these sensors may themselves be "smart", $i.e.$ have local processing. However, we will defer the critical sensor fusion functions to the compute capability of the MSSPs, dubbed "SmartInfra".

The SmartInfra has the following primary functions:

\subsubsection{Synthesis of Situational Awareness (SA) Information }

A key function of the SmartInfra would be to generate the SA information. 

It is to be noted that the MSSPs have ground-truth since they are stationary and can be precisely geo-located after installation. Therefore, GPS availability is not a concern at any time during the operation.
The combination of sensors such as LIDAR, RADAR, thermal and optical allow for 24-hour availability of information. 
The fusion of the sensors provides a rich starting point for further processing such as classification and subsequently contextualization, leading to good, first level SA information. This information can be further reconciled with observations from the different SCs engaged with the MSSP as well as neighboring MSSPs. To do this, model based “observers” of the environment shall be used to account for the dynamic movements of the various objects in the environment. Such environment observer calculations may be intensive and could partly be off-loaded to the cloud.

\subsubsection{SC Registration and Communication Management}

Another important function of the SmartInfra is to ensure appropriate handshake and communication with the individual SCs. Loosely speaking, each MSSP will function “like” a Cell Tower, and each SC will act “like” a cell phone. The SIETC gets broken down into “cells.” Continuing the analogy, a segment of the SIETC can be considered as a “city” with its own “Mobile Telephone Switching Office” operating from the cloud. 

The MSSP will then facilitate or manage the registration of the different SCs within its cell, and manage the communication by assigning appropriate communication bandwidth as needed.

\subsubsection{Incident Identification and Reporting}

As part of the information processing within an MSSP, in particular with the use of model-based “observers,” it would be possible to identify infrastructural and/or other issues (e.g., pot holes, accidents, stalled vehicles, etc.). The MSSP shall also identify such incidents and communicate to the authorities via level 3 communication through the cloud.

\subsection{Smart Connects (SCs)}

The SA information generated by the MSSPs need to be received by the different vehicles. This is done through the “SC device. The SC has three primary functions:

\subsubsection{Communication with MSSPs}

The SC will register itself into an SIETC through communication with the nearest MSSP. Then it will start sending information about its “Self-Awareness” (SeA) information to the MSSP, as well as receive the SA information from the MSSP.

\subsubsection{Decision Making}
The SC will then use the SA information to make decisions on behalf of the host of the SC. If the SC is hosted in an automobile, then it would decide on what would be the best tactical action to be taken by the vehicle (such as perform a lane change, slow down, accelerate, etc.). If the SC is hosted through a wearable device or smart phone, the SC would decide on the best indicators of information and diagnostics to be provided to the host.

\subsubsection{Decision Execution}

Once the SC makes a decision, the final function is to execute on the decision. When an SC is hosted on a wearable device, the execution is typically in the form of diagnostic warnings and messages.
When an SC is hosted on a DBW--enabled automobile, the SC will interface with the DBW system through well-defined APIs, in a secure fashion, to actually instruct the vehicle to perform maneuvers.

\subsection{Autonomous Driving with IEA}

The actual driving of the vehicle will be done by DBW capabilities that automotive OEMs will enable in their vehicles, with well-defined secure APIs that can be used by the SCs to define how the vehicle needs to be driven.

\section{A Framework for Assessing Benefits in the Proposed Architecture}

In this section, we will develop a mathematical framework that can be used to quantify the distribution of risk using the proposed architecture, relative to the existing paradigm for autonomous vehicles. Our overall rationale can be summarized as below:

\begin{enumerate}

\item Given that system level failures have always been happening historically, risk is essentially an estimate of how much aggregated "blame" will be assigned to one or more components for which a party takes responsibility.

\item The IEA defines  one functional decomposition of driving functionality and distributes the responsibility in alignment with the functional decomposition. This is leveraged to define the risk, for every system failure, as the aggregated "blame" for the individual driving function components.

\item Often multiple components fail simultaneously, and in such cases, we need a mechanism to define the blame for the individual components. Without any additional information, we will assume that each of the components at fault is equally to blame. We then can use the blame models similar to \cite{BlameModel}   

\item For any given system failure, the aggregated blame is the blame associated with every possible fault configuration that could result in the failure, weighted by the probability of that fault configuration.

\item To find the probability of the fault configurations we recognize that our system is a Bayesian Graph Network, and thereby leverage theoretical results therein.
\end{enumerate}

Let an autonomous vehicle system have $n$ components $C_1,C_2,\cdots, C_n$. A random variable $F_i$ is used to denote if the component $i$ is at fault or not. $F_i = 1$ indicates that the $i^{th}$ component is at fault and $F_i = 0$ indicates otherwise. We assume the random variables $F_1,\cdots, F_n$ are mutually independent. The fault configuration of the system is represented by $F=(F_1,F_2,\cdots,F_n)$. Let ${\mathcal{F}}$ denote the set of all the possible fault configurations, $i.e.$, $\mathcal{F}:=\{(f_1,\cdots,f_n): f_i\in \{0,1\}, i=1,\cdots,n\}$. The probability that the fault configuration F is equal to some $f\in \mathcal{F} $ is given by $P(F=f)$.\\

\begin{figure}[htb]
\centering{}
\includegraphics[scale=1]{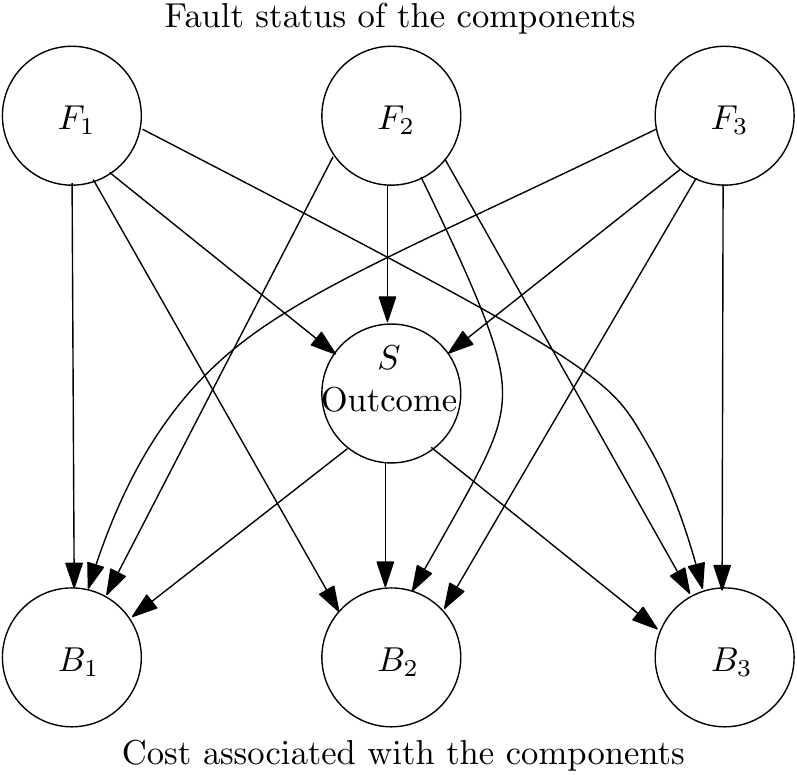}
\caption{A Bayesian network for the autonomous vehicle system with 3 components.}
\label{fig:bayes}
\end{figure}

Let $\mathcal{S}$ represent the set of all the possible outcomes of the functioning of the system. These outcomes are determined based on the severity and the risk level of the outcomes. Depending on the state of the components in the system, one can expect an outcome $S$ in $\mathcal{S}$. It is also assumed that the set of outcomes in $\mathcal{S}$ is mutually exclusive. Let the cost associated with the $i^{th}$ component be $B_i$, a discrete random variable. Let the set of all the possible values for $B_i$ be denoted as ${\mathcal{B}}$. If the outcome ($S=s$) and the fault states of all the components ($F=f$) are known, then $B_i:=\bar{B}_i(F=f,S=s)$ is a known value; however, as the cause of an outcome is only known through probabilities, $B_i$ is a random variable. The cost $B_i$ can be viewed as the "blame" or "responsibility" assigned to the $i^{th}$ component, and depends on the fault configuration and the outcome of the system. The random variables and their causal relationships is shown as a Bayesian network in Fig.\ref{fig:bayes} for a system with three components. This Bayesian network provides a model of the joint distribution of all the random variables in the system and their conditional dependencies.

The expected value of $B_i$ is then equal to $ \sum_{s\in {\mathcal{S}}} Exp(B_i/S=s) P(S=s)$. We now derive the expected value of $B_i$ given the outcome $S=s$.  
{\small
\begin{align}
&Exp(B_i/S=s)  = \sum_{y\in \mathcal{B}} y P(B_i=y/S=s) \nonumber \\
        & =  \sum_{y\in \mathcal{B}} y \sum_{f\in \mathcal{F}} P(B_i=y/F=f,S=s) P(F=f/S=s) \nonumber\\
        & =  \sum_{f\in \mathcal{F}} \sum_{y\in \mathcal{B}} y  P(B_i=y/F=f,S=s) P(F=f/S=s) \nonumber\\
         & =  \sum_{f\in \mathcal{F}} \bar{B}_i(F=f,S=s) P(F=f/S=s) \nonumber\\
         & =  \sum_{f\in \mathcal{F}} \bar{B}_i(F=f,S=s) \frac{P(S=s/F=f)P(F=f)}{P(S=s)} \nonumber \\
                & = \frac{1}{P(S=s)} \sum_{f\in \mathcal{F}} \bar{B}_i(F=f,S=s) P(S=s/F=f)P(F=f).
\end{align}
}

Based on a Hazards analysis ($e.g.$ see ISO26262 HAZOP),
we can infer the conditional probability, $P(S=s/F=f)$; that is, the likelihood of an outcome for a given fault configuration of the components. This is part of the standard process leading to ASIL levels for the component, which in turn leads to the level of scrutiny and verification performed
on the component. Note that in the above equations, $P(S=s)$ and $P(F=f)= \prod_{i} P(F_i=f_i)$, can be inferred from analysis of historical traffic and component data.

\subsection{Example}

To illustrate the computation of costs, consider the proposed architecture with three components: 1) Drive-By-Wire system, 2) Situational awareness generator, and 3) Decision making module. Suppose the cost of an outcome given the fault configuration and a scenario be defined as $ \bar{B}_i(F=f,S=s):=\frac{f_i}{\sum_i{f_i}} $. Let $P(F_i=1)= p_i$ for all $i=1,\cdots,3$. Let $ \mathcal{S}:= \{s_1,s_2,s_3,s_4\} $. The outcomes are organized sequentially in accordance to their severity levels such that $s_1$ represents an outcome with no accident and $s_4$ denotes an outcome with severe costs. $s_1$ happens only when no component is at fault, $i.e.$, $F=(0,0,0)$. In general, $s_i$ occurs if exactly $(i+1)$ components are at fault. \\

Let us compute $Exp(B_i/S=s_3)$ for $i=1,2,3$. 
\begin{align}
& Exp(B_1/S=s_3)  \nonumber \\ 
& = \frac{1}{P(S=s_3)}\sum_{f\in \mathcal{F}} \frac{f_1}{\sum_i{f_i}}{P(S=s_3/F=f)P(F=f)} \nonumber \\
& = \frac{1}{P(S=s_3)}\sum_{f\in \{(1,1,0),(1,0,1)\}} \frac{f_1}{\sum_i{f_i}}{P(S=s_3/F=f)P(F=f)} \nonumber \\
& = \frac{1}{P(S=s_3)}\sum_{f\in \{(1,1,0),(1,0,1)\}} \frac{1}{2}{P(F=f)} \nonumber \\
& = \frac{p_1(p_2(1-p_3) + (1-p_2)p_3) }{2P(S=s_3)}. \label{eq:1}
\end{align}
Similarly,
\begin{align}
Exp(B_2/S=s_3) & = \frac{p_2(p_1(1-p_3) + (1-p_1)p_3) }{2P(S=s_3)} \label{eq:2}
\end{align}

and, 

\begin{align}
Exp(B_3/S=s_3) & = \frac{p_3(p_1(1-p_2) + (1-p_1)p_2) }{2P(S=s_3)}. \label{eq:3}
\end{align}

Therefore, if the components are distributed, the cost or penalty incurred by each component is given by equations \eqref{eq:1}, \eqref{eq:2} and \eqref{eq:3} respectively. However, if all the components are controlled by a centralized entity, then the total cost incurred by this entity is given by:

\begin{align}
   & \sum_i Exp(B_i/S=s_3) \nonumber \\ & = \frac{p_1 + p_2 + p_3 + p_1p_2 + p_2p_3 + p_1p_3 - 3p_1p_2p_3}{2P(S=s_3)}.
\end{align}

Suppose $p_1=0.05$, $p_2 = 0.1$ and $p_3 =0.3$, then $Exp(B_1/S=s_3) \propto 17$, $Exp(B_2/S=s_3) \propto 32$ and $Exp(B_3/S=s_3) \propto 42$. Therefore, the proportion of responsibility assigned to the DBW system will be $100\times \frac{17}{91}\approx 18.6\%$. Similarly, the proportion of responsibility assigned to the Situational awareness generator will be $100\times \frac{32}{91} \approx 35.2\%$ and the proportion for the Decision making module will be $100\times \frac{42}{91} \approx 46.2\%$.

\section{Conclusions}

We have proposed a novel approach to accelerate the deployment of autonomous driving and correspondingly reap its benefits. Our concept is based on leveraging the explosive growth in “connectedness” and the possibilities it engenders. Specifically, we propose to reengineer the sensing and decision making of the autonomous car so that a significant portion of it is done external to the car in the infrastructure. First of all, the stationary nature of the infrastructure, including knowing the ground-truth about the location of all instruments, provides a superior situational awareness information to work with. More importantly, the proposed approach results in a fundamental re-distribution of the responsibilities and liabilities, that incentivizes the eco-system of businesses to accelerate the deployment of autonomous vehicles.

\subsection{Value Proposition}
If the IEA system could be implemented, there would benefits to several parties as below:
\begin{itemize}
    \item Automotive OEMs: OEMs stand to benefit the most from the availability of an IEA system, because this allows them to focus their efforts and energy on their core competencies and build safer cars. Most importantly, they will have a more manageable “liability.” 
    \begin{itemize}
        \item In the extreme scenario, OEMs do not need to add any sensors beyond what is available in production cars of today. On the other hand, they could continue to build navigation capabilities that could be used outside of the SIETCs and could be deployed at a pace with which they might be comfortable.
    \end{itemize}
    \item Infrastructure Operators: For infrastructure operators such as toll-booth operators as well as cell phone operators, the management of the infrastructure associated with the SIETCs would be an expansion of their current markets. the SIETCs create new markets and offer new opportunities for monetization of their services, breathing new life into businesses that are becoming more and more commoditized.
    \item Device Makers: The MSSP and the SCs become very rich business opportunities for entrepreneurs and businesses to begin manufacturing and installing on the infrastructure.
    \item SC Application Makers: The SC is not just used for driving the automobile. It can be used to communicate warnings and diagnostics to non-automated entities such as pedestrians and bicyclists and manually driven cars. This offers plenty of opportunities for entrepreneurs and businesses to come up with new applications – either working with the smartphones or the native SCs.
    \item Law enforcement, Infrastructure Maintenance, and Traffic Management: The presence of the infrastructure and sensing capabilities provides opportunities for newer applications leveraging the infrastructure to support traffic management, infrastructure maintenance, and law enforcement.
    \item Society at Large: The new paradigm will accelerate the penetration of automated driving overall, and correspondingly accelerate the reaping of the societal, environmental and economic benefits expected from autonomous vehicles in general.
\end{itemize}

This article has only proposed a concept for enabling autonomous driving. There are still many open research questions that need to be addressed in order to develop and realize this concept. But beyond the research, there are many questions that lead us to end with a note of caution: While the proposed concept is a powerful new way of looking at autonomous driving, we end this article with important questions that remain to be addressed before this concept will be feasible, as follows:
\begin{enumerate}
\item While we have developed a mathematical framework to show how the risk is reduced, will/can the liability be split in practice between the OEMs, infrastructure providers, and third-party companies?
\item How vulnerable will the IEA system be to cybersecurity issues?
\item Will the OEMs be willing to part with the collateral opportunities presented by going fully autonomous – specifically, the ability to acquire massive amounts of data that can be monetized on its own merit?
\item Will the OEMs be willing to work with each other to promote a common standard for communication with the infrastructure? This has been a challenge even with the simplified Basic Safety Messages (BSMs) for V2I and V2V initiatives.
\item 	Will the commercial infrastructure companies be willing to invest given the need to interface and liaison with a multitude of local government agencies in whose jurisdiction the infrastructure will lie?
\item	Can we clearly demonstrate that the technology will indeed be superior to existing purely autonomous vehicle technologies?
\end{enumerate}

\section*{Acknowledgment}

Useful discussions with Drs. Swaroop Darbha, Ray Korok, and Srikanth Saripalli of Texas A \& M University, and Drs. Ed Seymour, Ginger Goodin and Chris Poe of Texas A \& M Transportation Institute  are gratefully acknowledged.

\ifCLASSOPTIONcaptionsoff
  \newpage
\fi

\bibliographystyle{IEEEtran}
\bibliography{references}





\end{document}